\documentclass[a4paper,amsmath,amssymb,floatfix,prb,twocolumn,footinbib,superscriptaddress,preprintnumbers]{revtex4}
 
\usepackage{graphicx}    
\usepackage{dcolumn}     
\usepackage{bm}          
\usepackage{color}
\usepackage[colorlinks=true ,citecolor=blue]{hyperref}

\definecolor{mycolor}{rgb}{0,0.6,0}

\date{\today}

\begin{document}

\title{Pseudo-spin-dependent scattering in carbon nanotubes}
\author{Leonhard Mayrhofer}
\affiliation{Fraunhofer IWM, W\"ohlerstra\ss e 11, D-79108 Freiburg, Germany}
\author{Dario Bercioux}
\affiliation{Freiburg Institute for Advanced Studies, Albert-Ludwigs-Universit\"at, D-79104 Freiburg, Germany}
\affiliation{Physikalisches Institut, Albert-Ludwigs-Universit\"at, D-79104 Freiburg, Germany}

\begin{abstract}
The breaking of symmetry is the ground on which many physical phenomena are explained. This is important in particular for bipartite lattice structure as graphene and carbon nanotubes, where particle-hole and pseudo-spin are relevant symmetries. Here we investigate the
role played by the defect-induced breaking of these symmetries in the electronic scattering properties  of armchair single-walled carbon nanotubes. From Fourier transform of the local density of states we show that the active electron scattering channels depend on the conservation of the pseudo-spin. Further, we show that the lack of  particle-hole symmetry is responsible for the pseudo-spin selection rules observed in several experiments. This symmetry breaking arises from the lattice reconstruction appearing at defect sites. Our analysis gives an intuitive way to understand the scattering properties of carbon nanotubes, and can be employed for newly interpret several experiments on this subject. Further, it can be used to design devices such as pseudo-spin filter by opportune defect engineering.
\end{abstract}

\maketitle

\section{Introduction}
The unique electronic properties of single-walled carbon nanotubes (SWNTs) | related to their	 unusual	band structures~\cite{saito:1998,ando:2005}|have attracted great attention in fundamental and applied research because of the possibility of exploring phenomena unique to one-dimensional systems.~\cite{charlier:2007,postma:2001} Defectless SWNTs are well characterized both theoretically and experimentally | however, the presence of defects or tube endings is in fact crucial for the observation of quantum mechanical coherence phenomena.~\cite{vanema:1999,bockrath:2001,lemay:2001} In this respect, scanning tunneling microscopy/spectroscopy (STM/STS) represents a powerful instrument of investigations. This is an unparalleled tool able to measure the local electronic properties of SWNTs and correlate these with the atomic structure of the tube.  
Thanks to this technique, it has been possible to visualize Friedel oscillations~\cite{ouyang:2002:a,ouyang:2002:b,lee:2004} and quasi-bound-states~\cite{buchs:2009,bercioux:2010} in metallic SWNTs. These articles, reconstructing the low-energy scattering spectrum of SWNTs, have routinely shown the presence of an asymmetry in the scattering properties of left- and right-moving electrons. However, only in the work by Ouyang~\emph{et~al.}~\cite{ouyang:2002:b} there is an attempt to explain this asymmetry by relating it to the nature of defects. Their general conclusion is that this asymmetry roots in the different symmetry properties of the $\pi$ and $\pi^*$ bands in the armchair SWNT compared to the symmetry of defects.

%
%
\begin{figure}
	\centering
	\includegraphics[width=\columnwidth]{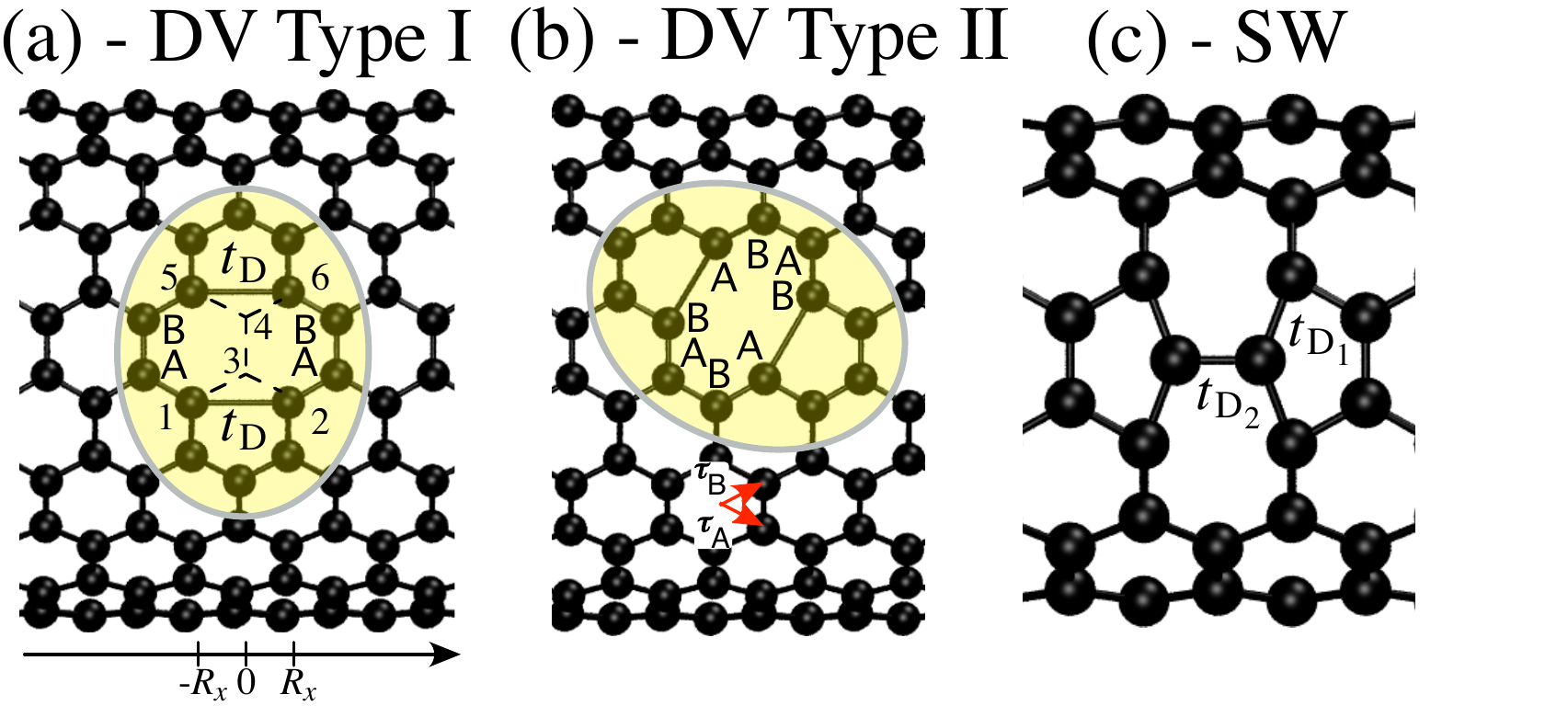}		
	\caption{\label{fig:one}  The DV defect in Panel (a) (Type I) is symmetric with respect to an exchange of the two carbon atoms A and B in each unit cell of the honeycomb lattice, whereas for the DV defect in Panel (b) (Type II) this symmetry is broken. In Panel (a), the atoms 3 and 4 are missing in the atomic structure while the bonds 1--2 and 5--6 represent the lattice reconstruction.  Panel (c) is a SW defect. (See main text for further details.)}
\end{figure}
%
%
Here, we present an in depth study of the scattering properties of electrons in armchair SWNTs in presence of structural defects. 
We show how the scattering properties are altered when fundamental symmetries of SWNTs, as the pseudo-spin symmetry and particle-hole symmetry (PHS), are broken by the defects. In particular we focus on  \emph{di-vacancy} (DV)  and Stone-Wales (SW) defects (c.f.~Fig.~\ref{fig:one}). These are  prevalently observed in graphene and SWNTs,~\cite{ando:2005,charlier:2007,hashimoto:2004,lee:2005,amorim:2007,berber:2008} however our approach is quite general and can be extended to other types of defects.   We show that pseudo-spin symmetry is determining the active scattering channels. Further, contrary to the hypothesis by Ouyang~\emph{et~al.}~\cite{ouyang:2002:b}, we substantiate that the experimentally observed asymmetry of electron scattering arises from the breaking of PHS caused by the reconstruction of the chemical bond between carbon atoms at the defect site. 
PHS breaking is also at the origin of other experimental observations~\cite{lee:2004,buchs:2009,bercioux:2010} unexplained so far. This newly revealed feature of defects in bipartite lattices can also be exploited to design a defect-based device able to filter the pseudo-spin species.

\section{Model and Formalisms}
%
%
\begin{figure}[t]
	\centering
	\includegraphics[width=\columnwidth]{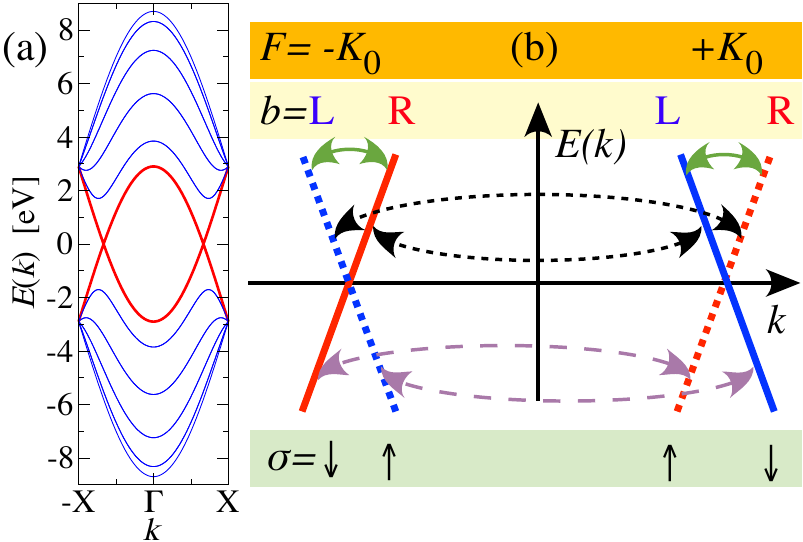}		
	\caption{\label{fig:two}   Complete energy spectrum of an armchair SWNT [Panel (a)] and its linearization around the charge neutrality point [Panel (b)]. In Panel (b), the full arrows represent IntraVB scattering, dashed arrows are InterVF scattering and dotted arrows are InterVB scattering. The dashed and solid lines represent electronic states with different pseudo-spin $\sigma$.}
\end{figure}
%

We consider an armchair SWNT in presence of DV and SW defects. 
Simulations~\cite{amorim:2007,berber:2008,lee:2005} and experiments~\cite{hashimoto:2004} indicate that DVs are likely the prevailing defect-type in SWNTs and graphene. Additionally, DV defects in SWNTs energetically favor the so-called 585 reconstruction~\cite{amorim:2007,berber:2008} where two new chemical bonds forms, therefore creating two pentagons and an octagon [c.f. Figs.~\ref{fig:one}(a) and \ref{fig:one}(b) | named here Type I and Type II, respectively]. SW defects consist of a local $\pi/2$ rotation of a C-C bond [c.f.~Fig.~\ref{fig:one}(c)], creating two pentagons and heptagons.
We describe the SWNT as a finite stripe of graphene with periodic boundary conditions along the boundaries delimited by the chiral vector of the tube. We treat the $\pi$-orbital electrons within a  tight-binding approximation. The defectless Hamiltonian with spin-independent hopping term $t$ reads $\mathcal{H}_0=-t \sum_{\langle i,j\rangle}  c_i^\dag c_j$, where the sum runs over all pairs of nearest-neighbor carbon atoms and $c_{i}^{\dagger}$ [$c_i$] is the creation [annihilation] operator for a $p_z$ electron at atom $i$. 
%
%
Two different methods are used in order to study  armchair SWNTs: an approximate low-energy one where only the linear dispersing modes are considered | named Method A (c.~f.~App.~\ref{app:a}), and a exact numerical method within tight-binding model | named Method B. Here, the SWNT is placed seamlessly between two semi-infinite leads at the same chemical potential consisting as well of SWNTs. Within Method A, we evaluate modifications to the scattering properties induced by defects via Fermi's Golden Rule~\cite{Bruus:2004} (c.~f.~App.~\ref{app:a}). With Method B, the local density of states (LDOS) $\mathcal{D}(x,E)$ is evaluated  as the imaginary part of the retarded Green's functions~\cite{datta:2005}  (c.~f.~App.~\ref{app:b}). Then for each energy value $E$ a Fourier transform~(FT) is performed on the LDOS for the coordinate along the tube axis in order to obtain the FT-LDOS $\mathcal{D}(2k,E)$. Here, the factor 2 originates from the fact that $|\psi(x)|^2$ is probed.

%
%

Within the tight-binding approximation the Hamiltonian $\mathcal{H}_\text{DV}$ describing a 585 DV  is obtained by removing the hopping terms to the missing carbon atoms and  by introducing hopping elements $t_\text{D}$ for the newly formed carbon bonds. 
It is important to note that these new bonds link carbon atoms of the same sub-lattice.  For the case of SW defects, two distinct reconstruction hopping elements $t_{\text{D}_1}$ and $t_{\text{D}_2}$ ($t_{\text{D}_1}<t<t_{\text{D}_2}$) are introduced into the Hamiltonian $\mathcal{H}_\text{SW}$.

In general a defect can give rise to three different types of scattering processes [c.~f.~Fig.~\ref{fig:two}(b)]: inter-valley forward/backward (InterVF/B) and intra-valley backward (IntraVB). However, only InterVB processes are pseudo-spin conserving.

In the following we consider the two possible orientations of a 585 DV defect on an armchair SWNT shown in Fig.~\ref{fig:one}(a)-(b), Type I and II, respectively,  and the SW defect in Fig.~\ref{fig:one}(c). 

\subsection{Type I di-vacancy defects}
\begin{figure}
	\centering
	\includegraphics[width=\columnwidth]{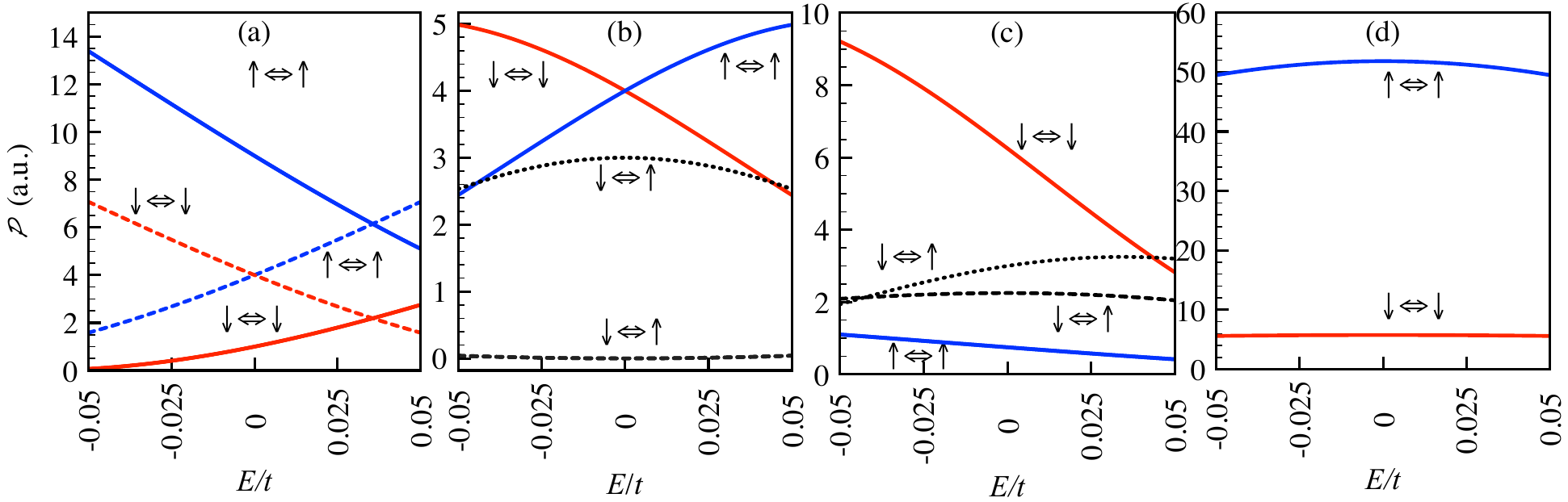}
	\caption{\label{fig:three}  Unnormalized scattering probabilities as a function of the energy for different scattering events. Panel (a) refers to Type I DV defect and Panels (b) and (c) to Type II DV defect. In all Panels red and blue lines refers to pseudo-spin  conserving processes of $\downarrow$ and $\uparrow$ carriers, respectively. In Panel (a) we show only InterVB  processes, where the dashed and the solid lines refer to $t_\text{D}=0$ and $t_\text{D}=0.5\,t$, respectively. In Panels (b) and (c) the solid  lines refer to InterVB scattering, the dashed-black lines to InterVF scattering and the dotted-black lines to IntraVB scattering. The lattice reconstruction parameter is $t_\text{D}=0$ in  Panel (b) and $t_\text{D}=0.5\,t$ in Panel (c).  The value of $t_\text{D}$ is obtained from DFT calculations~\cite{note:DFT}. In Panel (d) we display the case of the SW defect with $t_{\text{D}_1}=1.2t$ and $t_{\text{D}_1}=0.8t$. }
\end{figure}

The Hamiltonian associated with this defect commutes with the pseudo-spin symmetry operator $\mathcal{S}$ (c.~f.~App.~\ref{app:c1}), therefore the pseudo-spin $\sigma$ is a good quantum number. Within Method A, the scattering matrix element of the perturbation
$\mathcal{H}_\text{DV}$ between different SWNT eigenstates $\{|\sigma F \kappa\rangle\}$  reads
%
%
\begin{widetext}
\begin{align}\label{eq:four}
\langle\sigma F\kappa|\mathcal{H}_\text{DV}|\sigma'F'\kappa'\rangle =-\frac{1}{N_L} \Big\{ t_\text{D} \cos[(\Delta_+ +\delta_+)R_x] (1+\sigma\sigma')   + 2 t \cos\!\left[\frac{\Delta_+ +\delta_+}{2}R_x\right]\cos\!\left[\frac{\Delta_- +\delta_-}{2}R_x\right]\!(\sigma+\sigma')\Big\}.
\end{align}
\end{widetext}
%
%
Here $\Delta_\pm=F\pm F'$, $\delta_\pm=\kappa\pm\kappa'$, and $R_x=\sqrt{3}a_\text{CC}/2$. 
The conservation of the pseudo-spin $\sigma$  is also immediately evident from Eq.~(\ref{eq:four}) since only pseudo-spin conserving scattering processes ($\sigma = \sigma'$) have a non vanishing amplitude.  
Therefore, only InterVB processes are induces by  Type I DV defects.
Equation~(\ref{eq:four}) yields two different energy dependent scattering probabilities $\mathcal{P}(E)$ for states with pseudo-spins $\sigma = \pm = \uparrow/\downarrow$
%
%
\begin{align}\label{eq:five}
\mathcal{P}_{\sigma \Leftrightarrow \sigma }(E) & = \frac{2}{N_L^2} \left[  t_\text{D} - 2\,\sigma\, t \cos\left(\frac{2\pi}{3} +\sigma \frac{E R_x}{\gamma}\right)\right]^2 .
\end{align}
%
%
Here  $E = \pm \gamma \kappa$ with $\gamma = 3a_\text{CC}t/2 $ is the energy dispersion around  the charge neutrality point (CNP). 
Without lattice reconstruction ($t_\text{D}=0$)   the two scattering probabilities are particle-hole symmetric 
%
%
\[
\mathcal{P}_{\uparrow \Leftrightarrow \uparrow}(E) = \mathcal{P}_{\downarrow \Leftrightarrow \downarrow}(-E).
\]
%
%
However, the lattice reconstruction breaks PHS (c.~f.~App.~\ref{app:c2}). For $t_\text{D}\neq 0$ and energy values around the CNP, we observe that the InterVB scattering process $\downarrow\Leftrightarrow\downarrow$ is strongly suppressed compared to the $\uparrow\Leftrightarrow\uparrow$ process. This is illustrated in Fig.~\ref{fig:three}(a) where the full and the dashed lines refer to $t_\text{D}\neq0$ and $t_\text{D}=0$, respectively. 
Hence, we can clearly associate the different scattering probabilities to the breaking of PHS associated with the lattice reconstruction.

\begin{figure}[t]
	\begin{center}
	\includegraphics[width=\columnwidth]{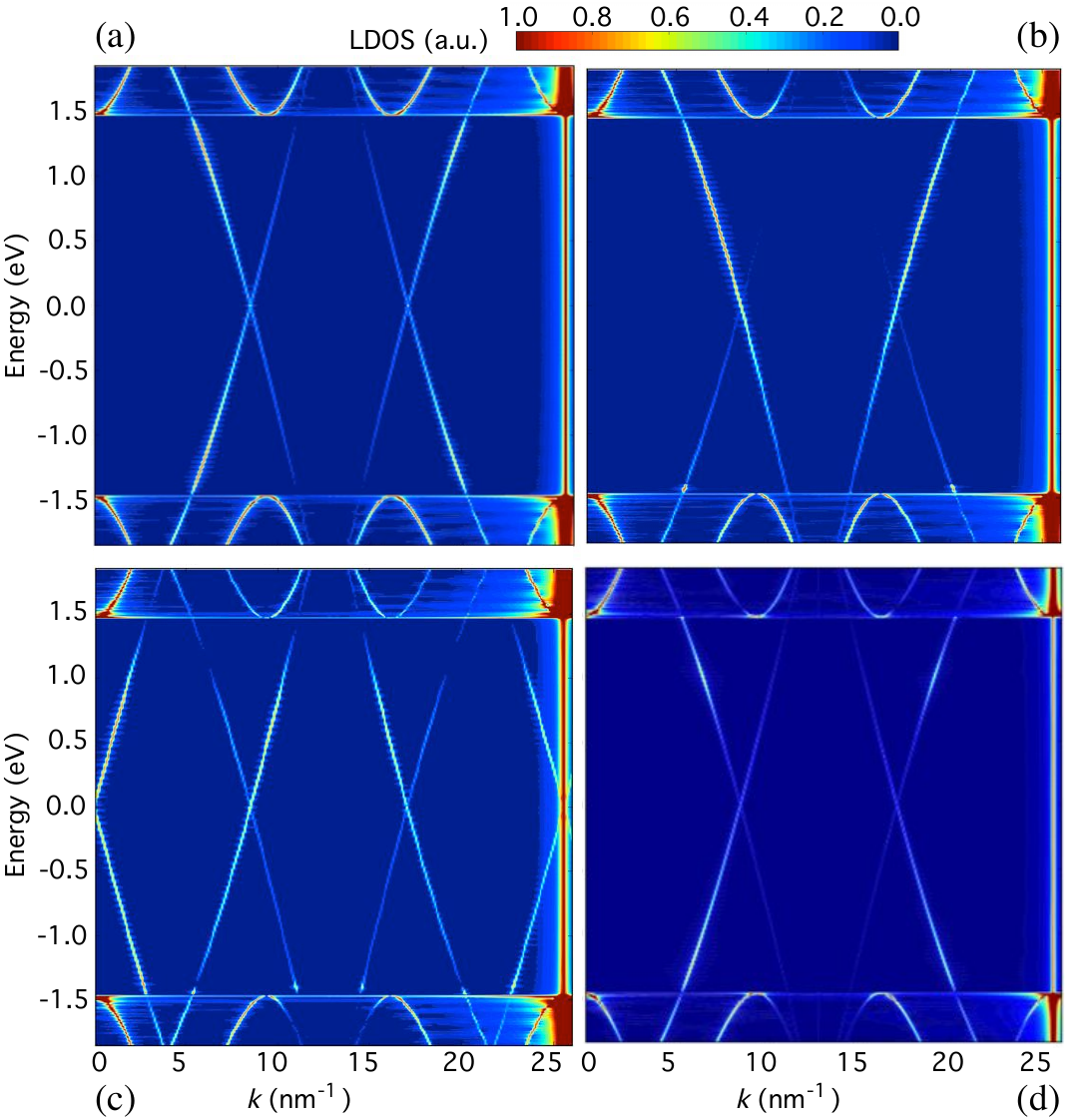}
	\caption{\label{fig:four}  Density plot of the Fourier transformed local density of states as a function of the exchanged momentum and energy for the case of a single Type I, Type II DV and a SW defect. In Panel (a) we have a Type I DV defect ($t_\text{D}=0$) whereas in Panel (b) $t_\text{D}=0.5t$.  In Panel (c) we have a Type II DV defect with $t_\text{D}=0.5\ t$. In Panel (d) we have a SW defect with $t_{\text{D}_1}=1.2t$ and $t_{\text{D}_1}=0.8t$. Note that for energies higher than $\pm|1.5|$~eV the nonlinear SWNT-subbands appear.}
	\end{center}
\end{figure}

Using Method B, we consider a (5,5) armchair SWNT with a single Type I DV defect. In Fig.~\ref{fig:four} we display the FT-LDOS $\mathcal{D}(2k,E)$ for the case without [panel (a)] and with lattice reconstruction $t_\text{D}=0.5t$ [panel~(b)]. For energies within the range $\pm1.5$~eV we find a direct image of the linear SWNT spectrum around $E=0$. This signature corresponds to InterVB scattering as explained by the \emph{extended} $\bm{k}\cdot\bm{p}$ model~\cite{buchs:2009,bercioux:2010}. 
In the absence of lattice reconstruction the FT-LDOS is particle-hole symmetric [Fig.~\ref{fig:four}(a)], 
%
%
\[
\mathcal{D}(2k,E)=\mathcal{D}(2k,-E)
\]
%
%
whereas the lattice reconstruction introduces a strongly reduced scattering probability for the $\sigma = \downarrow$ scattering channel [c.f. Fig.~\ref{fig:four}(b)] compared to the $\sigma = \uparrow$ scattering channel around the CNP.
%
%
%
%
%
%
%
%
%

\subsection{Type II di-vancancy defects}
The Hamiltonian associated with the tilted  Type II defect does not commute with the pseudo-spin symmetry operator $\mathcal{S}$ in contrast to the  Type I defect.
In this case we obtain more intricate results for the scattering processes. Method A shows that the pseudo-spin non-conserving InterVF and IntraVB processes have a finite probability [c.f.~Figs.~\ref{fig:three}(b) and \ref{fig:three}(c)]. Again, without lattice reconstruction the scattering is particle-hole symmetric, \emph{i.e.}, 
%
%
\begin{align*}
\mathcal{P}_{\uparrow \Leftrightarrow \uparrow}(E) & = \mathcal{P}_{\downarrow \Leftrightarrow \downarrow}(-E) \\ 
\mathcal{P}_{\sigma \Leftrightarrow -\sigma}(E)& = \mathcal{P}_{\sigma \Leftrightarrow -\sigma}(-E)
\end{align*}
%
%
 hold [c.f. Fig.~\ref{fig:three}(b)]. However for $t_\text{D}\neq 0$  PHS is broken again | we observe, contrary to the previous case, a strong enhancement of the scattering probability for the  $\downarrow\Leftrightarrow\downarrow$ InterVB processes whereas the $\uparrow\Leftrightarrow\uparrow$ InterVB scattering is almost suppressed [c.f. Fig~\ref{fig:three}(c)]. This is confirmed also by Method B. In Fig.~\ref{fig:four}(c) we show the FT-LDOS for the case of a Type II DV defect  with a finite value for the lattice reconstruction ($t_\text{D}=0.5t$). Compared to the case of Type I DV defect are now clear the IntraVB scattering processes at $k\gtrsim 0~\text{nm}^{-1}$. 

\subsection{Stone-Wales defects} 
The SW defects intrinsically include  a lattice reconstruction | therefore  PHS is  broken by its very nature. For an infinite SWNT there is a degeneracy for the SW defect related to clockwise and anti-clockwise rotations of the C-C bond. In order to correctly describe  these defects, we introduce a symmetrized Hamiltonian accounting for both configurations. This Hamiltonian commutes with the pseudo-spin symmetry operator $\mathcal{S}$, therefore also in this case the pseudo-spin $\sigma$ is a good quantum number. Indeed,
we find by using  Method A that also for the SW Hamiltonian only  scattering processes conserving the pseudo-spin are allowed [Fig.~\ref{fig:three}(d)]. 

In Fig.~\ref{fig:four}(d) we present the case of a single SW defect in a (5,5) SWNT within  Method B. Here again there are no signs of processes where pseudo-spin  is not conserved. However, opposite to the DV cases, the $\mathcal{D}(2k,E)$ shows a more complex dependance on the energy and pseudo-spin.

\section{Case of two di-vacancies}
\begin{figure}[t]
	\begin{center}
	\includegraphics[width=\columnwidth]{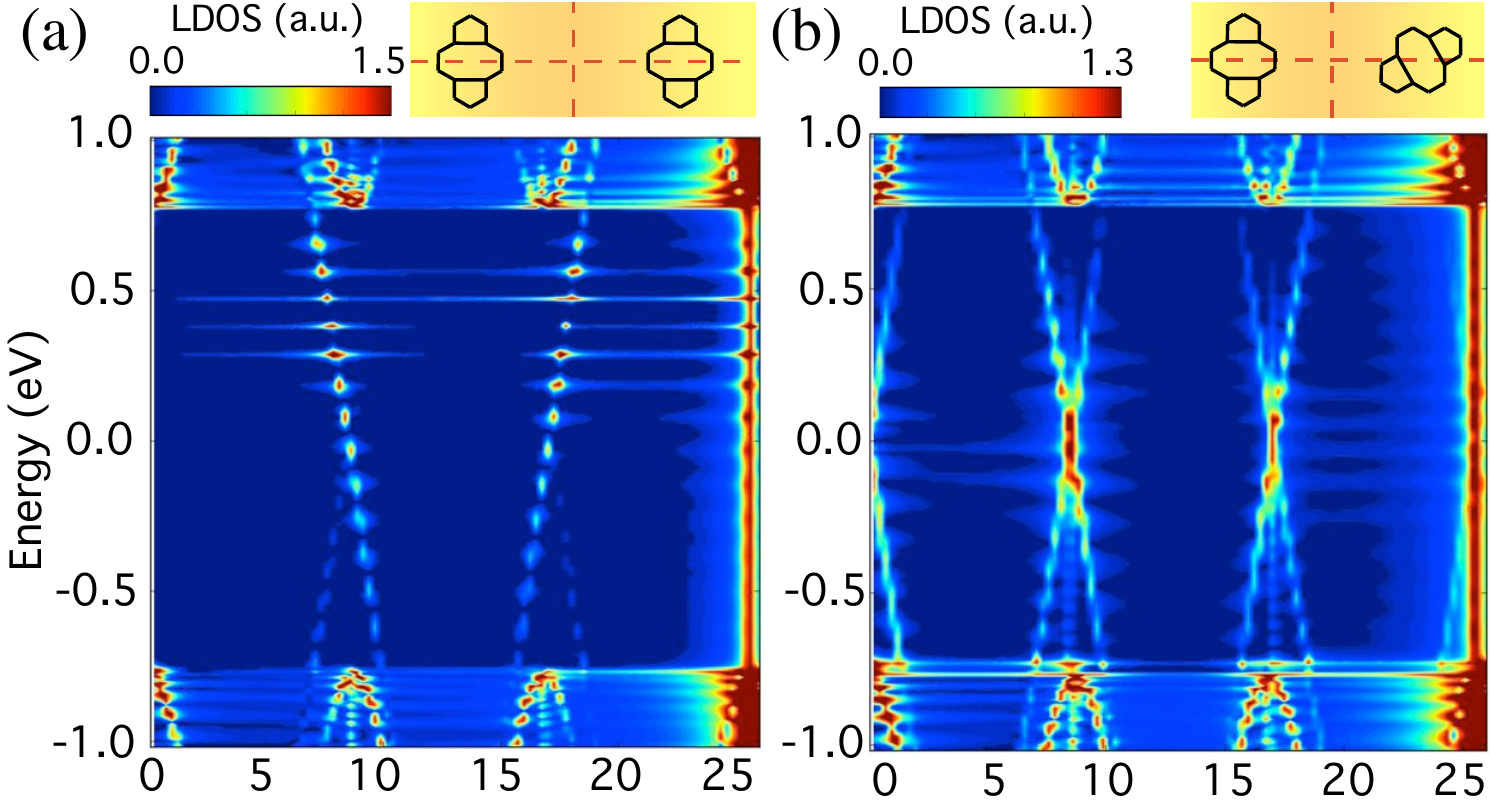}
	\caption{\label{fig:five}  Density plot of the Fourier transformed local density of states as a function of the exchanged momentum and energy for the case of two 585 DV defects with $t_\text{D}=0.5t$. In Panel (a) we consider two Type I DV defects , in Panel (b) two Type I and Type II DV defects. The signal at $k\sim25.5$~nm$^{-1}$ is associated with $R_x$, the smallest length scale of the armchair SWNT along the $x$-axis.}
	\end{center}
\end{figure}

We investigate | only within Method B | a (10,10) armchair SWNT with two reconstructed Type I and II DV defects ($t_\text{D}=0.5t$) placed at a distance of $L= 14.5$~nm along the SWNT axis. Since $t_\text{D}\neq 0$, PHS is already broken. Four unequal combinations can be formed from the two  Type I and II defects | however we consider here only two relevant cases as depicted in the insets of Fig.~\ref{fig:five}. The scattering behavior of each of the single defects determines the pattern of $\mathcal{D}(2k,E)$ between the defects. The overall features of the FT-LDOS for the combination of two Type I DV  defects [Fig.~\ref{fig:five}(a)] resembles those of a  single Type I DV  defect shown in Fig.~\ref{fig:four}(b). We observe, however, a discretization in energy indicating the formation of quasi-bound-states between the defects. The pseudo-spin-selection rule for Type I DV  defects accounts for the fact that those bound states are essentially only formed by $\sigma=\uparrow$ states  as the asymmetry in the amplitude between the different linear slopes clearly shows.  Figure~\ref{fig:five}(b) contains  one Type II DV  defect that breaks the pseudo-spin symmetry|now InterVF and IntraVB scattering occur. The fingerprint of the InterVF processes are the vertical lines at  $k\sim 9~\text{nm}^{-1}$ and $\sim 17~\text{nm}^{-1}$. The IntraVB scattering is revealed by the dispersive features at $k\gtrsim 0~\text{nm}^{-1}$.

\section{Theory \emph{vs.}\ Experiments}

We now apply these findings to interpret the experimental results of Ouyang~\emph{et~al.}~\cite{ouyang:2002:a,ouyang:2002:b} and Lee~\emph{et~al.}~\cite{lee:2004}. In the first work the scattering at a single isolated defect in an armchair SWNT is analyzed. The reconstructed energy spectrum shows a selection rule of InterVB scattering processes with respect to the pseudo-spin. In view of our results, we can  attribute the asymmetry to the lattice reconstruction. In the second case the standing waves of electrons scattered at one end of a nanotube peapod are examined. Also in this case a clear asymmetry in the scattering processes is found. We can relate  this asymmetry to the peapod at the tube end. There various hexagons of the lattice structure are replaced by pentagons therefore mixing the two sub-lattices of the SWNT as in the case of the DV and SW defects. Thus the breaking of PHS explains the asymmetry observed experimentally. In the experiments by Buchs~\emph{et~al.}~\cite{buchs:2009} the quasi-bound-states in defected metallic SWNT were investigated. There the defects were constituted mainly by DVs created by means of ions irradiation~\cite{bercioux:2010}. In order to make a comparison we have analyzed the case of two DV defects in two possible spatial configurations. Also in this case we have observed an absence of processes that are not conserving the pseudo-spin in the case of two defects Type I. For the second configurations where processes not conserving pseudo-spin are allowed, the numerical outcome accounts quite well for the experimental results.

%
%
%
%
Summarizing these finding we recognized two main features affecting the electron scattering at defects in the lattice structure of an armchair SWNT: the conservation of the pseudo-spin degree-of-freedom, and the breaking of PHS due to the lattice reconstruction. The former determines the active scattering channels, the latter introduces an asymmetry among the different scattering processes.

\section{Perspective} 

The methods we have introduced here can also be employed to study defects in graphene and chiral SWNTs. In the latter case, the  pseudo-spin is in general not conserved. However, the breaking of PHS due to the lattice reconstruction is a general feature of the underlying SWNT honeycomb lattice and does not depend on the tube chirality.

Choosing the symmetry of the defect accordingly, it is possible to modify the ratio of the transmission probabilities of the two pseudo-spin species. Therefore, the prospect of engineering definite defects in SWNTs, would permit to filter electrons with a specific degree-of-freedom | the pseudo-spin | in analogy to a similar effect on valley-spin induced by line defects in graphene~\cite{gunlycke:2011}.

 As we have shown with our results, the pseudo-spin dependent scattering response  changes considerably  with the electron energy.  
 Moreover, different defects can be combined in a line, so as we have demonstrated, in order to obtain an enhanced effect of filtering.
 Therefore, by an opportune defect sequencing and by fixing the appropriate energy it should be possible to select the desired pseudo-spin in a controlled way. The problem of engineering well defined defects, \emph{e.g.} via ion irradiation, can be overcome by employing adatoms instead of DV and SW defects. These can be placed on the SWNTs with an atomic precision using, \emph{e.g.} the tip of a STM. In fact, adatoms  break PHS in the same manner as the DV and SW defects by allowing hopping between $\pi$ electrons of carbon atoms of the same sub-lattice\cite{lim:2007}.

\appendix
\section{Method A}\label{app:a}
In proximity of the charge neutrality point the energy spectrum of armchair SWNTs can be approximated by two linearly dispersing branches centered around the two valleys $K_0=\pm4\pi/3\sqrt{3}a_\text{CC}$, where $a_\text{CC}\approx0.142$~nm is the C-C bond length (c.f. Fig.~\ref{fig:two}). In this approximation the electronic wave function can be expressed as \cite{mayrhofer:2008}
%
%
\begin{equation}\label{eq:one}
\langle \bm{r}| \sigma F\kappa \rangle=\sum_{p=\text{A,B}}f_{p\sigma}\varphi_{pF\kappa}(\bm{r})\,.
\end{equation}
%
%
Here $F=\pm K_0$ is the valley index, $\kappa$ is the momentum relative to $F$, and $\sigma$ is the  pseudo-spin. 
This is defined by $\sigma=\text{sgn}(bF)$, where $b=\pm=\text{Right/Left}$ is the electron motion direction.
The coefficients $f_{p\sigma}$ account for the two inequivalent carbon atoms ($p=\text{A,B}$) in the lattice structure. These are defined as
%
%
\begin{equation}\label{eq:two}
f_{p\sigma}=\frac{1}{\sqrt{2}}\begin{cases}
1 & p= \text{A} \\
\sigma & p=\text{B}
\end{cases}\,.
\end{equation}
%
%
The function $\varphi_{pF\kappa}(\bm{r})$ describes the armchair SWNT wave function component on sub-lattice $p$.
The functions $\varphi_{pF\kappa}(\bm{r})$ in Eq.~\eqref{eq:one} is defined as 
%
%
\[
\varphi_{pF\kappa}(\bm{r})=\frac{1}{\sqrt{N_L}}\sum_{\bm{r}_p} \text{e}^{\text{i} (F + \kappa )R_x} \chi(\bm{r}-\bm{r}_p),
\]
%
%
where the sum runs over the SWNT lattice sites, $2N_L$ is the total number of carbon atoms, and $\chi(\bm{r}-\bm{r}_p)$ are the localized $p_z$ orbitals at the positions $\bm{r}_p=\bm{R}+\bm{\tau}_p$ on sub-lattice $p$, and  $\bm{\tau}_p$  the displacement vector within the honeycomb lattice unit cell~\cite{mayrhofer:2008}. Finally, $R_x$ is the projection of the lattice vector onto the tube axis (c.f.~Fig.~\ref{fig:one}a).

Within  Fermi's Golden Rule approximation, the transition probability between different states $\{|\sigma F \kappa\rangle\}$ is given by
%
%
\[
\mathcal{P}_{f\leftarrow i}\propto \left|{}_f\langle \sigma F \kappa|\mathcal{H}_\text{defect}|\sigma F \kappa\rangle_i\right|^2 \delta(E_f-E_i)
\]
%
%
where $E_{i/f}$ are the energies of the initial and final states, respectively.

\section{Method B}\label{app:b}
The local density of states (LDOS) has been evaluated numerically via the retarded Green's functions 
%
%
\begin{equation}
\mathcal{G}(E)=\frac{1}{E-\mathcal{H}+\Sigma_\text{R} + \Sigma_\text{L}}.
\end{equation}
%
%
Here $\mathcal{H}$ is the Hamiltonian of the defected SWNT, and  $\Sigma_\text{R/L}$ are the self-energies of the right or left lead, respectively. They have been determined by a recursive Green's function method.~\cite{datta:2005} From $\mathcal{G}(E)$ we have evaluated the LDOS $\mathcal{D}(x,E)$ at carbon atom site $x$ via
%
%
\begin{equation}
\mathcal{D}(x,E)=-\frac{1}{\pi}\text{Im}\left[\left\langle x\left|\mathcal{G}(E)\right|x\right\rangle\right]
\end{equation}
%
%
where $\text{Im}[\ldots]$ is the imaginary part.
In order to get a smooth LDOS on the surface of the SWNT, we have performed a convolution of $\mathcal{D}(x,E)$ with a function of the form $\exp(-\lambda r)$ where $r$ is the distance of a fictitious tip of a STM from the carbon atom at $x$, and $\lambda$ is an opportune constant.  

\section{The symmetries}\label{app:c}
\subsection{The pseudo-spin symmetry}\label{app:c1}
In our case the pseudo-spin operator $\mathcal{S}$ is given by the Pauli matrix $\sigma_x$. In general, electron scattering is pseudo-spin conserving if the defect Hamiltonian commutes with $\mathcal{S}$. Since the effect of $\mathcal{S}$ is only a simultaneous exchange of the two carbon atoms of the unit cell, see Fig.~\ref{fig:one}(a), pseudo-spin is conserved if and only if the defect geometry is symmetric with respect to an exchange of the A and B atoms within the single unit cells. Hence it is clear that the Hamiltonian associated with the Type I DV defect commutes with $\mathcal{S}$, whereas this is not true for the case of the Type II DV defect. In the case of the symmetrized version of the SW defect in Fig.~\ref{fig:one}(c), its Hamiltonian commutes with the pseudo-spin operator.

\subsection{The particle-hole symmetry}\label{app:c2}
The particle-hole symmetry operator for the SWNT is defined as $\mathcal{U}=-\sigma_z$.
A Hamiltonian $\mathcal{H}$ is particle-hole symmetric if the following properties hold: $|p[h]\rangle=\mathcal{U}|h[p]\rangle$ where $|p\rangle[|h\rangle]$ is a particle[hole] state, and $\{\mathcal{H},\mathcal{U}\}=0$. When the lattice reconstruction is bonding together carbon atoms of the same sub-lattices, the Hamiltonian associated with the reconstructed defect is not anti-commuting with $\mathcal{U}$.

\begin{acknowledgments} 
\noindent We thank G.~Buchs, H.~Grabert, L. Lenz, C. Lieber, O.~Gr\"{o}ning,  and M.~Moseler for useful discussions. The work of DB is supported by the DFG grant BE~4564/1-1 and by the Excellence Initiative of the German Federal and State Governments.  
\end{acknowledgments}


\begin{thebibliography}{99}

\bibitem{saito:1998} R. Saito, G. Dresselhaus, and M. Dresselhaus, \emph{Physical Properties of Carbon Nanotubes} (Imperial College Press, London, 1998).

\bibitem{ando:2005} T. Ando, J. Phys. Soc. Jpn. \textbf{74}, 777 (2005).

\bibitem{charlier:2007} J.-C. Charlier, X. Blase, and S. Roche, Rev. Mod. Phys. \textbf{79}, 677 (2007).

\bibitem{postma:2001}  H.~W.~C. Postma,  T. Teepen, Z. Yao, M. Grifoni, and C. Dekker, Science \textbf{293}, 76 (2001).

\bibitem{vanema:1999}  L. C. Vanema,  J. W. G.  Wild\"oer, J. W.  Janssen, S. J.  Tans, H. L. J. Temminck Tuinstra, L. P. Kouwenhoven, and  C. Dekke, Science \textbf{283}, 52 (1999).
 
\bibitem{bockrath:2001} M. Bockrath, W. Liang, D. Bozovic, J. H. Hafner, C. M. Lieber, M. Tinkham, and H. Park,  Science \textbf{291}, 283 (2001).

\bibitem{lemay:2001} S. G. Lemay, J. W.  Janssen, M.  van den Hout, M.  Mooij, M. J.  Bronikowski, P. A. Willis, R. E.  Smalley,  L. P. Kouwenhoven and C. Dekker, Nature \textbf{412}, 617 (2001).

\bibitem{ouyang:2002:a} M .Ouyang, J.-L. Huang, and C. M.  Lieber,  Annu. Rev. Phys. Chem. \textbf{53}, 201 (2002).

\bibitem{ouyang:2002:b} M .Ouyang, J.-L. Huang, and C. M.  Lieber, Phys. Rev. Lett. \textbf{88}, 066804 (2002).

\bibitem{lee:2004}  J. Lee, S. Eggert, H. Kim, S. J. Kahng, H. Shinohara, and Y. Kuk, Phys Rev Lett. \textbf{93}, 166403 (2004).

\bibitem{buchs:2009} G. Buchs, D. Bercioux, P. Ruffieux, P. Gr\"oning, H. Grabert, and O. Gr\"oning,  Phys. Rev. Lett. \textbf{102}, 245505 (2009).

\bibitem{bercioux:2010} D. Bercioux, G. Buchs, H. Grabert, and O. Gr\"{o}ning, Phys. Rev. B \textbf{83}, 165439 (2011).

\bibitem{hashimoto:2004} A. Hashimoto, K. Suenaga, A. Gloter, K. Urita, and S. Iijima, Nature \textbf{430}, 870 (2004).

\bibitem{lee:2005} G. D. Lee, C. Y.  Wang, E.  Yoon, N. M.  Hwang, D. Y. Kim, ans K. M.  Ho, Phys. Rev. Lett. \textbf{95}, 205501 (2005).

\bibitem{amorim:2007} R. G. Amorim, A. Fazzio, A. Antonelli, F. D. Novaes, and A. J. R. Da Silva,  Nano Lett. \textbf{7}, 2459 (2007).

\bibitem{berber:2008} S. Berber and A. Oshiyama,  Phys. Rev. B \textbf{77}, 165405 (2008).

\bibitem{mayrhofer:2008} L. Mayrhofer and M. Grifoni,  Eur. Phys. J. B \textbf{63}, 43 (2008).

\bibitem{Bruus:2004} H. Bruus, and K. Flensberg, \emph{Many-Body Quantum Theory in Condensed Matter Physics: An Introduction}
(Oxford University Press, USA , 2004).

\bibitem{note:DFT} L. Mayrhofer L. and D. Bercioux, \emph{in preparation}.

\bibitem{datta:2005} S. Datta, \emph{Quantum Transport: Atom to Transistor} (Cambridge University Press, Cambridge, 2005).

\bibitem{gunlycke:2011} D. Gunlycke and C. T.  White, Phys. Rev. Lett. \textbf{106}, 136806 (2011).

\bibitem{lim:2007} S. H. Lim, R. Li, W.  Ji, and J. Lin, Phys. Rev. B \textbf{76}, 195406 (2007).
\end{thebibliography}
\end{document}